\documentclass[aps,reprint,amssymb,prd,floatfix,superscriptaddress]{revtex4-1}

\usepackage[english]{babel}
\usepackage[utf8]{inputenc}
\usepackage[colorinlistoftodos]{todonotes}
\usepackage{neutron}
\usepackage{mathtools}

\begin{document}

\title{First direct constraints on Fierz interference in free  neutron $\bm\beta$ decay}


\author{K.~P.~Hickerson}		\affiliation{Kellogg Radiation Laboratory, California Institute of Technology, Pasadena, California 91125, USA} 
\author{X.~Sun}			\affiliation{Kellogg Radiation Laboratory, California Institute of Technology, Pasadena, California 91125, USA} 

\author{Y.~Bagdasarova}		\affiliation{Los Alamos National Laboratory, Los Alamos, New Mexico 87545, USA} \affiliation{Department of Physics, University of Washington, Seattle, Washington 98195, USA} 
\author{D.~Bravo-Berguño}	\affiliation{Department of Physics, Virginia Tech, Blacksburg, Virginia 24061, USA} 
\author{L.~J.~Broussard}	\thanks{Currently at Oak Ridge National Laboratory, Oak Ridge, TN 37831, USA}\affiliation{Los Alamos National Laboratory, Los Alamos, New Mexico 87545, USA} 
\author{M.~A.-P.~Brown}			\affiliation{Department of Physics and Astronomy, University of Kentucky, Lexington, Kentucky 40506, USA} 
\author{R.~Carr}			\affiliation{Kellogg Radiation Laboratory, California Institute of Technology, Pasadena, California 91125, USA} 
\author{S.~Currie}			\affiliation{Los Alamos National Laboratory, Los Alamos, New Mexico 87545, USA} 
\author{X.~Ding}			\affiliation{Department of Physics, Virginia Tech, Blacksburg, Virginia 24061, USA} 
\author{B.~W.~Filippone}		\affiliation{Kellogg Radiation Laboratory, California Institute of Technology, Pasadena, California 91125, USA} 
\author{A.~Garc\'ia}			\affiliation{Department of Physics, University of Washington, Seattle, Washington 98195, USA} 
\author{P.~Geltenbort}		\affiliation{Institut Laue-Langevin, 38042 Grenoble Cedex 9, France} 
\author{J.~Hoagland}		\affiliation{Department of Physics, North Carolina State University, Raleigh, North Carolina 27695, USA} 
\author{A.~T.~Holley}		\thanks{Currently at Dept. of Physics, Tennessee Tech University, Cookeville, TN, USA}\affiliation{Department of Physics, North Carolina State University, Raleigh, North Carolina 27695, USA} \affiliation{Department of Physics, Indiana University, Bloomington, Indiana 47408, USA} 
\author{R.~Hong}			\affiliation{Department of Physics, University of Washington, Seattle, Washington 98195, USA} 
\author{T.~M.~Ito}			\affiliation{Los Alamos National Laboratory, Los Alamos, New Mexico 87545, USA} 
\author{A.~Knecht}			\thanks{Currently at Paul Scherrer Institute, Villigen, Switzerland}
\affiliation{Department of Physics, University of Washington, Seattle, Washington 98195, USA} 
\author{C.-Y.~Liu}			\affiliation{Department of Physics, Indiana University, Bloomington, Indiana 47408, USA} 
\author{J.~L.~Liu}			\affiliation{Kellogg Radiation Laboratory, California Institute of Technology, Pasadena, California 91125, USA}  \affiliation{Department of Physics, Shanghai Jiao Tong University, Shanghai 200240, China}
\author{M.~Makela}			\affiliation{Los Alamos National Laboratory, Los Alamos, New Mexico 87545, USA} 
\author{R.~R.~Mammei}		\affiliation{Department of Physics, Virginia Tech, Blacksburg, Virginia 24061, USA} 
\author{J.~W.~Martin}		\affiliation{Department of Physics, University of Winnipeg, Winnipeg, MB R3B 2E9, Canada} 
\author{D.~Melconian}		\affiliation{Cyclotron Institute, Texas A\&M University, College Station, Texas 77843, USA} 
\author{M.~P.~Mendenhall}	\thanks{Currently at Physical and Life Sciences Directorate, Lawrence Livermore National Laboratory, Livermore, CA 94550, USA} \affiliation{Kellogg Radiation Laboratory, California Institute of Technology, Pasadena, California 91125, USA} 
\author{S.~D.~Moore}		\affiliation{Department of Physics, North Carolina State University, Raleigh, North Carolina 27695, USA} 
\author{C.~L.~Morris}		\affiliation{Los Alamos National Laboratory, Los Alamos, New Mexico 87545, USA} 
\author{R.~W.~Pattie,~Jr.}		
	\affiliation{Los Alamos National Laboratory, Los Alamos, New Mexico 87545, USA} 
\author{A.~P\'erez Galv\'an}	\thanks{Vertex Pharmaceuticals, 11010 Torreyana Rd., San Diego, CA 92121, USA} \affiliation{Kellogg Radiation Laboratory, California Institute of Technology, Pasadena, California 91125, USA}  
\author{R.~Picker}			\thanks{Currently at Physical Sciences Division, TRIUMF Laboratory, Vancouver, BC V6T 2A3, Canada}\affiliation{Kellogg Radiation Laboratory, California Institute of Technology, Pasadena, California 91125, USA} 
\author{M.~L.~Pitt}			\affiliation{Department of Physics, Virginia Tech, Blacksburg, Virginia 24061, USA} 
\author{B.~Plaster}			\affiliation{Department of Physics and Astronomy, University of Kentucky, Lexington, Kentucky 40506, USA} 
\author{J.~C.~Ramsey}		\affiliation{Los Alamos National Laboratory, Los Alamos, New Mexico 87545, USA} 
\author{R.~Rios}			\affiliation{Los Alamos National Laboratory, Los Alamos, New Mexico 87545, USA} \affiliation{Department of Physics, Idaho State University, Pocatello, Idaho 83209, USA} 
\author{A.~Saunders}		\affiliation{Los Alamos National Laboratory, Los Alamos, New Mexico 87545, USA} 
\author{S.~J.~Seestrom}		\affiliation{Los Alamos National Laboratory, Los Alamos, New Mexico 87545, USA}
\author{E.~I.~Sharapov}		\affiliation{Joint Institute for Nuclear Research, 141980, Dubna, Russia} 
\author{W.~E.~Sondheim}	\affiliation{Los Alamos National Laboratory, Los Alamos, New Mexico 87545, USA} 
\author{E.~Tatar}			\affiliation{Department of Physics, Idaho State University, Pocatello, Idaho 83209, USA} 
\author{R.~B.~Vogelaar}		\affiliation{Department of Physics, Virginia Tech, Blacksburg, Virginia 24061, USA} 
\author{B.~VornDick}		\affiliation{Department of Physics, North Carolina State University, Raleigh, North Carolina 27695, USA} 
\author{C.~Wrede}			\thanks{Currently at Department of Physics and Astronomy and National Superconducting Cyclotron Laboratory, Michigan State University, East Lansing, Michigan 48824, USA} \affiliation{Department of Physics, University of Washington, Seattle, Washington 98195, USA}  
\author{A.~R.~Young}		\affiliation{Department of Physics, North Carolina State University, Raleigh, North Carolina 27695, USA} 
\author{B.~A.~Zeck}			\affiliation{Department of Physics, North Carolina State University, Raleigh, North Carolina 27695, USA} \affiliation{Los Alamos National Laboratory, Los Alamos, New Mexico 87545, USA} 

\collaboration
{UCNA Collaboration}
\noaffiliation

\date{\today}

\begin{abstract}
Precision measurements of free neutron $\beta$-decay have been used to precisely constrain our understanding of the weak interaction.  
However the neutron Fierz interference term $b_n$, which is particularly sensitive to Beyond-Standard-Model tensor currents at the TeV scale, has thus far eluded measurement. Here we report the first direct constraints on this term, finding $b_n = 0.067 \pm 0.005_{\text{stat}} {}^{+0.090}_{- 0.061}{}_{\text{sys}}$, consistent with the Standard Model. The uncertainty is dominated by absolute energy reconstruction and the linearity of the beta spectrometer energy response.
\end{abstract}

\maketitle

Precision measurements in nuclear beta decay including lifetimes, angular/spin correlations and energy spectra can be used to test predictions of the electroweak sector of the Standard Model \cite{Erler2004,Severijns2006,Severijns_ann_rev2011,Bhattacharya2012,Gonzalez_N-C2013,
Young2014mxa,Baessler2014}. 
Current efforts are underway to measure many of these quantities in nuclear beta decay as well as in free neutron decay. 
The Fierz interference term 
vanishes in the Standard Model but serves as a probe for new physics in scalar and tensor couplings \cite{Erler2004, Bhattacharya2012}. 
In this paper, a direct measurement 
of the Fierz interference term for the free neutron 
(here denoted $b_n$) 
is presented for the first time. This term is particularly sensitive to tensor couplings that previous extractions of the Fierz term in superallowed $0^+\to 0^+$ beta decay are not. 

The Fierz interference term appears in the full form of the differential neutron decay rate parameterized
in terms of correlation coefficients of neutron spin,  $\vec{\sigma}_n=\vec{J_n}/|\vec{J_n}|$, and
momenta, $\vec{p_e},\vec{p_\nu}$, and total energies, $E_e,E_\nu$, of the final state particles
\cite{Jackson1957a}
\begin{align}
\label{eq: alphabet}
\begin{split}
	d\Gamma = 
	\mathcal{W}(E_e)
	\Bigl[
		1 
		&+ a \, \frac{\vec{p}_e \cdot \vec{p}_\nu}{E_e E_\nu} 
		+ b_n \frac{m_e}{E_e} 
		+ A \frac{\vec{p}_e \cdot \vec{\sigma}_n}{E_e} \\
		&+ B \, \frac{\vec{p}_\nu \cdot \vec{\sigma}_n}{E_\nu}
        + \cdots ~
	\Bigr] ~ dE_e dE_\nu\,d\Omega_e d\Omega_\nu
\end{split}
\end{align}
\noindent where $\mathcal{W}(E_e)$ includes the total decay rate (e.g. $1/\tau_n$) and the phase space along with recoil-order, radiative and Coulomb corrections. 
The correlation coefficients also include recoil-order corrections. 
The dimensionless $b_n$, is the only spin and momentum-direction independent coefficient, 
and thus survives summation over spin and integration over the final state angular distributions leaving a distribution dependent only on the electron energy, $E$:
\begin{equation}
\label{eq:bsm:Fierz-rate}
	d\Gamma_b(E) = \left(1 + b_n \frac{m_e}{E}\right)\mathcal{W}(E)~dE
\end{equation}
For the neutron, a combination of both Fermi and Gamow-Teller components, $\bF$ and $\bGT$ respectively, contributes to the Fierz term (see Refs \cite{Bhattacharya2012,Hickerson2013}):
\begin{equation}
\label{eq: bsm:neutron-Fierz}
	b_n =  \frac{\bF + 3 \lambda^2 \bGT}{1 + 3 \lambda^2}.
\end{equation}
where $\lambda \equiv \gA/\gV$ is the ratio of the axial vector to vector nucleon coupling constants. Note that the Fermi and Gamow-Teller components can also be described in terms of an effective field theory (EFT) framework \cite{Bhattacharya2012,Cirigliano2013} that relates new scalar and tensor quark level couplings to $\bF$ and $\bGT$ respectively.

The best limits for $\bF$ are from a global fit to multiple superallowed $J^\pi = 0^+\to0^+$ $\beta$-decay $ft$ values.
Hardy and Towner \cite{Hardy2014} place this limit at $\bF = -0.0028 \pm 0.0026$, or $|\bF| < 0.0043$ at 90\% C.L.  
Several $\beta$ decay experiments have set limits on $\bGT$ using the influence of $b$ on the correlation parameters of Eq. \eqref{eq: alphabet} including ${}^{19}$Ne \cite{Holstein1977}, ${}^{60}$Co \cite{Wauters2010}, ${}^{114}$In \cite{Wauters2009}, ${}^{67}$Cu \cite{Soti_67Cu2014}, and neutrino mass constraints \cite{Ito2005}, which give limits in the range $|\bGT|< 0.04 \text{---} 0.13$ at 90\% confidence level.
These limits are derived from the consistency of the observed correlation coefficient with the Standard Model prediction, assuming that $\bGT$ is the dominant Beyond-Standard Model (BSM) contribution. 
Reviews of limits on tensor contributions in nuclear $\beta$-decay can be found in \cite{Severijns2006,Severijns_ann_rev2011,Gonzalez_N-C2013}. 
In the EFT approach \cite{Bhattacharya2012,Cirigliano2013} sensitive limits to new tensor couplings, including $\bGT$, can be obtained also from pion decay \cite{PiBeta2009}. 
It has also been pointed out \cite{Bhattacharya2012,Gonzalez_N-C2013,Cirigliano2013} that, while nuclear beta and pion decays are found to provide the strongest constraints, at present, to BSM tensor couplings to left-handed neutrinos, measurements at the LHC provide the best constraints for tensor couplings to right-handed neutrinos.

For the free neutron, similar limits have been obtained by using the precision correlation parameter $A$ \cite{Pattie-Hickerson2013,Wauters-Garcia2014}.
While no measurements of the direct spectral extraction of the Fierz interference term have been published, several precision measurements are underway using $^6$He \cite{6He_Huyan2016,6HeKnecht2013} and the free neutron \cite{PocanicNAB2008}. 
The sensitivity of these searches is discussed in \cite{G-A_N-C_kinematic}. 
Extraction of the Fierz term from measurements of neutron decay has the advantage of the well-understood theoretical treatment of the decay, eliminating the need for nuclear structure corrections which complicate the interpretation of some nuclear decays.
In particular, recent reviews such as ref \cite{Gonzalez_N-C2013}, for example, do not include some (nominally quite stringent) tensor limits from suppressed nuclear decays in which nuclear structure effects hinder the decay relative to strongly allowed decays.
Because of its unique characteristics, the UCNA experiment described below, provides an opportunity to directly measure $b_n$ and thus $\bGT$ with comparable precision to nuclear decays. 

\label{section:UCNA}
The ultracold neutron asymmetry (UCNA) experiment
is the first experiment\cite{WPattieJr2008} to use ultracold neutrons (UCN)
in a precision measurement of neutron decay correlations.
The 2010 data set from the UCNA experiment provides a precision measurement of the $\beta$ asymmetry parameter, $A$, with a fractional error $<$ 1\% \cite{Mendenhall2013}. 
Because of the 4$\pi$ acceptance for the decay electrons, the very low ambient and neutron-generated backgrounds (Signal:Noise $>$ 120:1), and energy reconstruction at the 1\% level, this data set also provides a precision measurement of the $\beta$ decay spectrum. This allows, for the first time, a direct spectral extraction of the Fierz interference term, $b_n$, for the free neutron. 
Experience with high precision beta decay spectroscopy during the intensive search for a neutrino with 17 keV rest mass for example, highlighted the need for a detailed and quantitative analysis of scattering and energy loss effects for these measurements \cite{Wietfeldt:1995ja}. Detailed models of the UCNA instrument response over the past ten years of operation allow these effects to be evaluated experimentally and provide a firm foundation for the evaluation of sources of systematic uncertainty in this experiment. 

Details of the UCNA experiment are discussed in \cite{Plaster2012,Mendenhall2013,Hickerson2013,Mendenhall2014}; here the components of the experiment that allow a measurement of the $\beta$ spectrum are described.
\begin{figure}
	\includegraphics[width=3.5in]{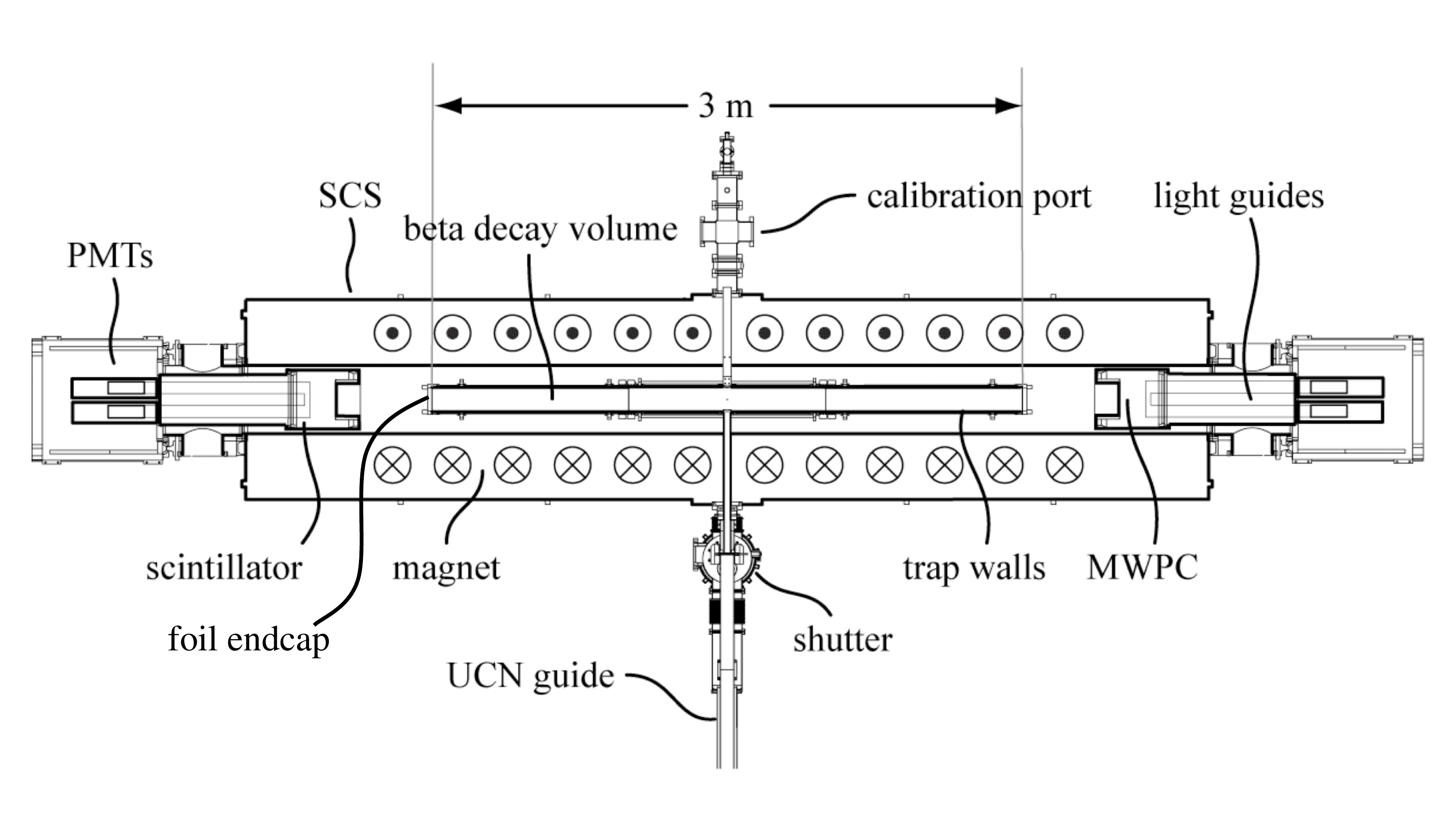}
	\caption{Schematic diagram of the UCNA spectrometer.}
			 \label{fig:scs-section-detail}
\end{figure}
A schematic diagram of the UCNA spectrometer is shown in Fig. \ref{fig:scs-section-detail}. UCN, generated by the UCN source at the Los Alamos Neutron Science Center \cite{Saunders2003,Morris2002}, are polarized by a 7T superconducting magnet and transported to the copper decay trap centered in the superconducting spectrometer \cite{PlasterSCS2008} that provides a 1T spin-holding field. 
UCN are confined by 0.7 $\mu$m thick beryllium-coated mylar foils at the ends of the copper tube decay chamber, trapping UCN while allowing decay electrons to pass through and be transported, via the magnetic field of the spectrometer, to two detectors, located on either end of the decay trap.

Each detector is composed of a multi-wire proportional chamber (MWPC) \cite{ItoMWPC2007} and a 3.5 mm thick plastic scintillator to measure the electron energies. Scintillation light from each detector is directed, via light guides, towards four Photo-Multiplier Tubes (PMTs). Gain stabilization is achieved using individual $^{207}$Bi sources embedded in small scintillator blocks attached to each PMT \cite{Mendenhall2013,Morris1976}. Event triggers require signals above threshold in at least two PMTs and valid events also require a signal in the MWPC. The position measurement done by the MWPC allows a calibration of the position-dependent energy response of the scintillator system (e.g. due to  optical transport in the scintillator and light guides) which  improves the electron energy reconstruction. This position dependent response is measured by loading the spectrometer with neutron-activated xenon. By observing the decay spectrum features (mainly the 915 keV endpoint from $^{135}$Xe) as a function of position using the MWPC, the position-dependent light transport of the scintillators is determined. 
The energy response and linearity of each PMT is calibrated
with conversion electron sources ($^{139}$Ce,$^{113}$Sn, and $^{207}$Bi) inserted horizontally, transverse to the spectrometer axis, at different locations across the center of the decay trap. Energy loss due to the sealing foils of each source is determined using a collimated $^{241}$Am alpha source and a silicon detector. Overall, the energy response has a low energy threshold of $\simeq$ 60 keV, an energy resolution of $7\%/\sqrt{E}$ (where $E$ is the electron kinetic energy in MeV) dominated by photo-electron statistics and a linearity of $\simeq$ 1\%. The uncertainty in the absolute electron kinetic energy varies from $\pm$2.5 keV at 130 keV to $\pm$6.5 keV at 1 MeV. 




\label{section:super-sum}

For the extraction of the $\beta$ decay asymmetry in the UCNA experiment, a super-ratio \cite{Plaster2012,Mendenhall2013} is formed in order to cancel, to first order, differences in detection efficiency between the two detectors as well as spin-dependent loading efficiencies of the UCN. Because the UCNA data is taken with polarized neutrons, the $\beta$ decay spectrum includes an energy dependent modification of the spectrum due to the presence of the asymmetry term, $A$ in Eq. (\ref{eq: alphabet}). To remove this dependence from the spectral analysis a ``super-sum'' is introduced, described below.

In the UCNA apparatus, four detector count rates are measured corresponding to the two detectors and the two neutron spin directions. 
These rates can be written as a function of decay electron total energy ($E$) and angle ($\theta$) between the neutron spin and electron momentum by using Eq. (\ref{eq: alphabet}) and integrating over the neutrino momentum:
\begin{equation}
	\begin{aligned}
	\label{eq:rate-model}
    r^\uparrow_1(E) &= \tfrac{1}{2} \eta_1 
		N^\uparrow\left ( 1 + b_n m_e/E + Ay(E)\right )
			 \, \mathcal{W}(E),
	\\
    r^\uparrow_2(E) &= \tfrac{1}{2} \eta_2 
		N^\uparrow \left (1 + b_n m_e/E - Ay(E)\right )
           	\, \mathcal{W}(E),
    \\
    r^\downarrow_1(E) &= \tfrac{1}{2} \eta_1 
		N^\downarrow \left ( 1 + b_n m_e/E - Ay(E)\right )
           \, \mathcal{W}(E),
	\\
    r^\downarrow_2(E) &= \tfrac{1}{2} \eta_2 
		N^\downarrow \left ( 1 + b_n m_e/E + Ay(E)\right )
            \, \mathcal{W}(E),
	\end{aligned}
\end{equation}
where e.g. $r^\uparrow_2$ corresponds to the rate in detector 2 for spin $\uparrow$, 
    \, 
    $y(E) \equiv \expected{P} \beta \expected{\cos \theta}$,
with $\expected{P}$ the average polarization, and $\beta = v/c$.
These four rates are expressed in terms of the detector efficiencies,
$\eta_{1,2}(E)$. 
The UCN loading numbers for each spin state, $N^\uparrow$ and $N^\downarrow$, 
differ typically by 50\% due to polarized UCN transport through the magnetic fields of the polarizing magnet and the spectrometer magnet.

An electron energy spectrum that does not have 
a significant dependence on $A$ can be generated [up to $\cal{O}$ $(b_n A^2$)] by forming a super-sum as the 
sum of the geometric means of the spin/detector pairs:
\begin{equation}
	\label{eq:super-sum}
    \Sigma(E) \equiv 
		\tfrac{1}{2} \sqrt{ r^\uparrow_1 r^\downarrow_2 }
      + \tfrac{1}{2} \sqrt{ r^\uparrow_2 r^\downarrow_1 },
\end{equation}
where, using the rates given in Eq. (\ref{eq:rate-model}),
\begin{equation}
    \Sigma(E) = {
		\sqrt{\eta_1 \eta_2 N^\uparrow N^\downarrow } (1 + b_n m_e/E) \mathcal{W}(E)
	}.
\end{equation}
While this does not eliminate the detector efficiencies, 
it does remove dependence on $A$ from the extraction of $b_n$.

For extraction of $b_n$ and for analysis of potential systematic uncertainties, a GEANT4 simulation \cite{Geant4} described in \cite{Mendenhall2013,Mendenhall2014}, has been modified to include the Fierz term and to incorporate a newer version of GEANT4 (Geant4.10.2).  


In the simulation, the 1 T magnetic field of the spectrometer directs electrons towards the detectors on either end, where energy loss in the UCN trap windows, the MWPC and its windows and the plastic scintillator are calculated. The detectors are located in a field expansion region of 0.6T to suppress electron backscattering. A post processor is then used to convert the energy loss into photons, including light quenching via Birk's Law \cite{BirksLaw} that was previously calibrated for the UCNA scintillator \cite{Yuan2006}. This result is then converted into a PMT signal including the effects of energy resolution, due to PMT response shot noise, and low-energy threshold effects.

Two types of initial energy distributions are used for the simulation described above: a pure Standard Model (i.e. $b_n = 0$ in Eq. (\ref{eq:bsm:Fierz-rate})) and a maximal Fierz distribution in which 
\begin{equation}
\label{eq:bsm:Fierz-rate2}
	d\Gamma_b(E) = \left(\frac{m_e}{E}\right)\mathcal{W}(E)~dE.
\end{equation}
In both cases the recoil-order, radiative and Coulomb corrections are included (see \cite{Mendenhall2014}). Note that these corrections produce a small $m_e/E$ term of order $1\times 10^{-3}$ \cite{Bhattacharya2012}.
The Fierz term is then extracted by fitting the experimental decay spectrum, $\Sigma(E)$, to a superposition of the Monte Carlo generated Standard Model super-sum and the maximal 
Fierz super-sum with $b_n$ as a free parameter. 


The direct, spectral measurement of $b_n$ is essentially a measurement of a small distortion of the energy spectrum compared to the Standard Model distribution. For example, with a simple allowed phase space spectrum, a $b_n = 0.1$ corresponds to a global shift in the peak of the neutron decay spectrum downward by $\simeq 5$ keV. Thus small uncertainties in the energy response (specifically the absolute energy and linearity) can lead to significant systematic uncertainties in the extraction of $b_n$. 

In contrast, statistical uncertainties can be quite small for a large data sample. The statistical uncertainties can be estimated using \cite{Gluck1995} where, assuming $b_n\simeq 0$ and using the full energy spectrum, $\sigma_b = 7.5/\sqrt{N}$ where $N$ is the number of detected events.  Because of detection energy threshold effects, fitting the spectrum at low energies can be problematic. 
Using a restricted energy range as in \cite{Gluck1995}, and with energy dependent detector efficiency, 
\begin{equation}
	\sigma_b^{-2} = m_e^2 N \left(\expected{E^{-2}} - \expected{E^{-1}}^2\right)
\end{equation}
which, for the electron kinetic energy window used in this work from 150 keV to 650 keV, gives $\sigma_b = 11.4/\sqrt{N}$. Even for this range, with $2.0\times 10^7$ events from the 2010 UCNA data set, the corresponding statistical uncertainty is $\sigma_b < 0.003$, which is much smaller than the systematic uncertainties as described below. 

An initial assessment of systematic effects was performed using an analytical model in place of a full simulation \cite{Hickerson2013}. Here the electron energy is generated from the allowed phase space and a model of detector response is used to account for detector efficiency, electron backscattering, background subtraction and energy response effects including energy resolution, non-linearities and absolute energy calibration. These studies indicated relatively modest systematic uncertainties ($\sigma_b$) from background subtractions ($\pm 0.005$), energy resolution ($\pm 0.01$) and electron backscattering from spectrometer windows and detectors ($<\pm 0.005$). The backscattering effects are minimized by using events that triggered only one scintillator and its adjacent MWPC. The energy dependence of the detector efficiency is estimated to give $\sigma_b \leq0.02$, assuming an uncertainty of $\pm 20\%$ in the calculated inefficiency. 
The detector inefficiency due to energy deposition in material along the electron beam path (e.g. detector windows) is determined from the GEANT4 Monte Carlo. The PMT threshold response is determined from the data using overdetermined triggers, since a trigger requires only two PMTs above threshold. We note that the efficiency is $>90\%$ above the minimum energy used for the analysis and the simulated energy deposition is expected to be well-reproduced by the GEANT4 simulation (see \cite{Hardy_scin_resp2008}). However these analytical studies suggest considerably larger uncertainties ($\ge\pm 0.05$) from non-linearities and absolute energy calibration. 

\begin{figure}
    \includegraphics[width=\linewidth]{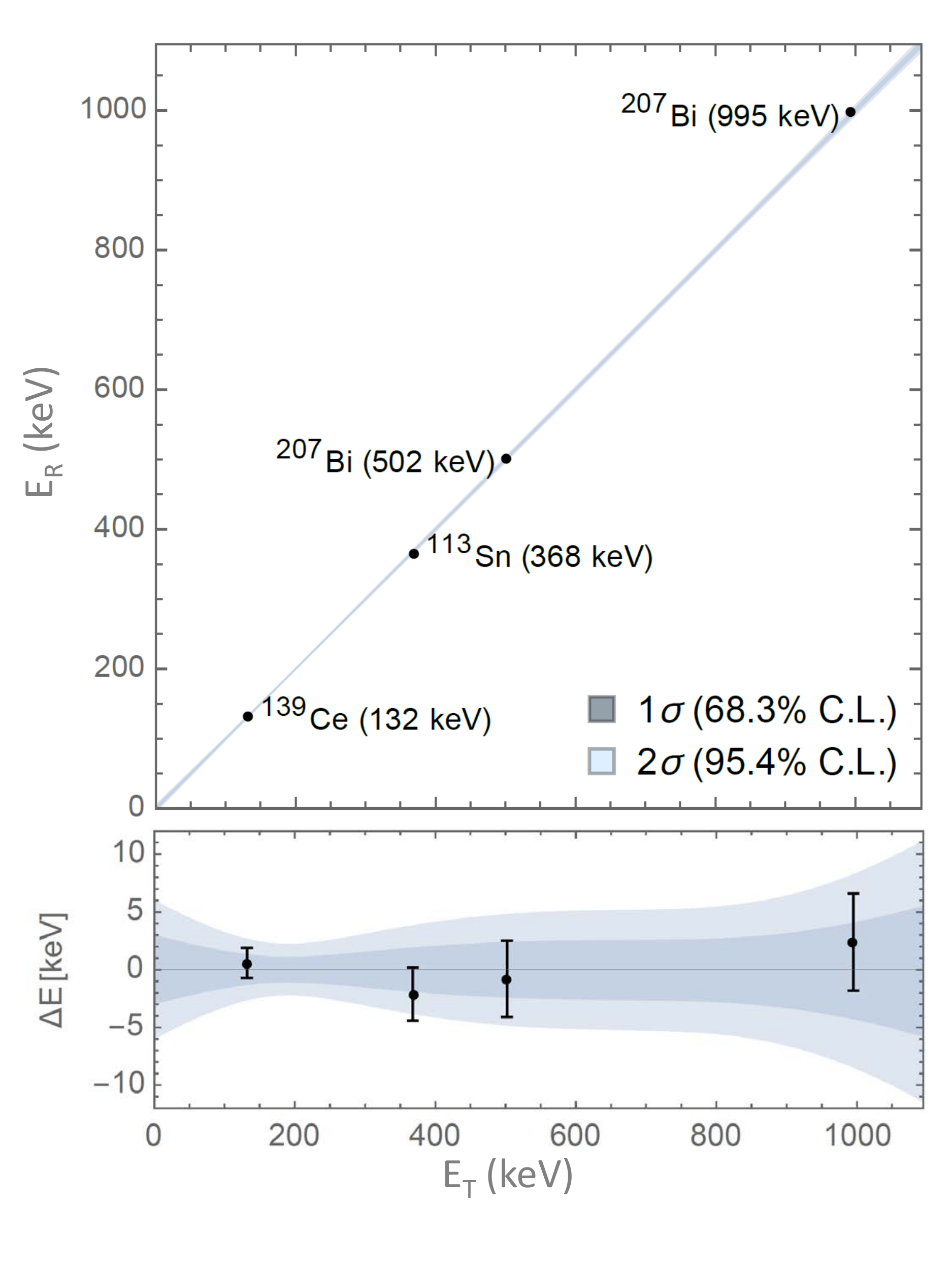}
\vskip -.3in
 \caption{Top: Reconstructed kinetic energy vs. true kinetic energy for the conversion electron calibration sources. Uncertainty bands for the quadratic energy response functions used in the Monte Carlo estimate of systematic uncertainty in $b_n$ are also shown. Bottom: $\Delta E$, the difference between true energy and reconstructed energy vs. true energy. The bands are the same as for the top plot. The mean energies of conversion electron sources and their $1\sigma$ uncertainty are also shown.} 
    \label{fig:source-uncertainty}
\end{figure}

To better quantify the uncertainty due to energy response, the full GEANT4 simulation of the spectrometer is used to investigate how the uncertainty in energy response could contribute to a false $b_n$. 
As discussed above, this reconstructed kinetic energy response, $E_R$, is determined from a series of calibration runs with conversion electron sources. These sources have an
approximately mono-energetic conversion electron with true kinetic energy $E_T$, determined by averaging the individual electron lines over the resolution of the detectors. However this calibration has a corresponding uncertainty \cite{Mendenhall2014} due to variations in the detector position response, uncertainties in gain stabilization, etc. These uncertainties are indicated in Fig. \ref{fig:source-uncertainty} for the four energies used in the calibration from the three conversion electron sources. 
The $E_R$ determined from the calibrations assumes a linear response for conversion of energy deposition to light output after correction for light quenching discussed above. The assumption of linear response is confirmed in Fig. \ref{fig:source-uncertainty} as the observed $\Delta E=E_T-E_R$ is consistent with zero within its uncertainty. These uncertainties are the standard deviation of many global fits to the energy response based on 10-12 separate source location runs taken during each of five separate time periods spread throughout the experiment.
\begin{figure}
  \includegraphics[width=\linewidth]{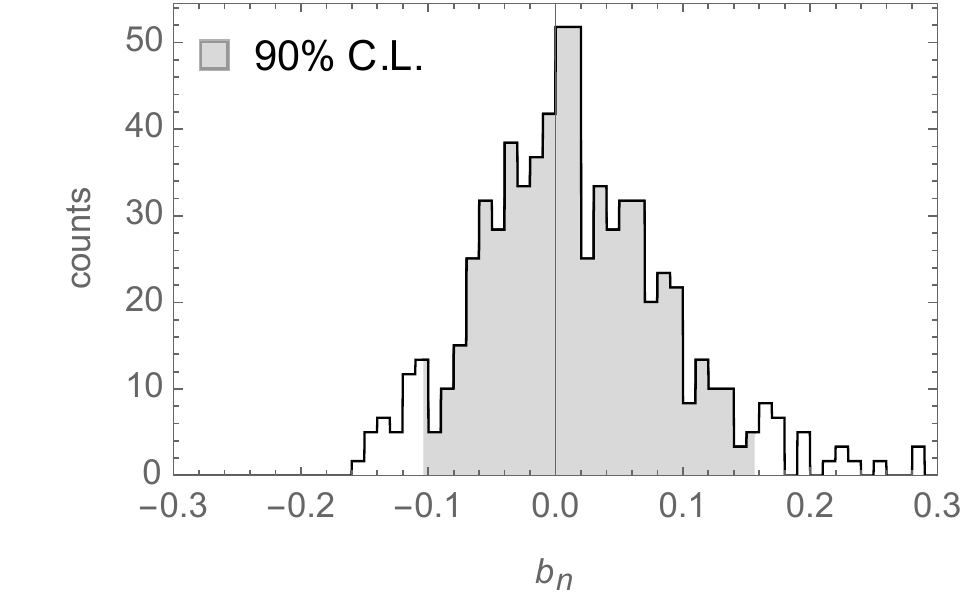}
  \caption[b_nshistogram]
			{Distribution of fitted values of $b_n$ for the simulations with adjusted energy response.}
            \label{fig:bn-histogram}
\end{figure}

To quantify the systematic uncertainty in $b_n$ due to energy response, Monte Carlo simulations with $b_n=0$ are performed where the energy response is varied with variations consistent within the uncertainties of $\Delta E$. To do this, the $E_R$ is assumed to be a non-linear polynomial as a function of $E_T$. The coefficients for this polynomial are then sampled in order to reproduce the calibration uncertainties (assumed Gaussian). Both quadratic and cubic polynomials were sampled in separate studies (higher orders are not justified due to the limited impact on $b_n$), but the results are not significantly different. The envelopes for these sampled energy responses for a quadratic polynomial (both $1\sigma$ and $2\sigma$) are shown as bands in Fig. \ref{fig:source-uncertainty}.
\begin{figure}
  \includegraphics[width=\linewidth]{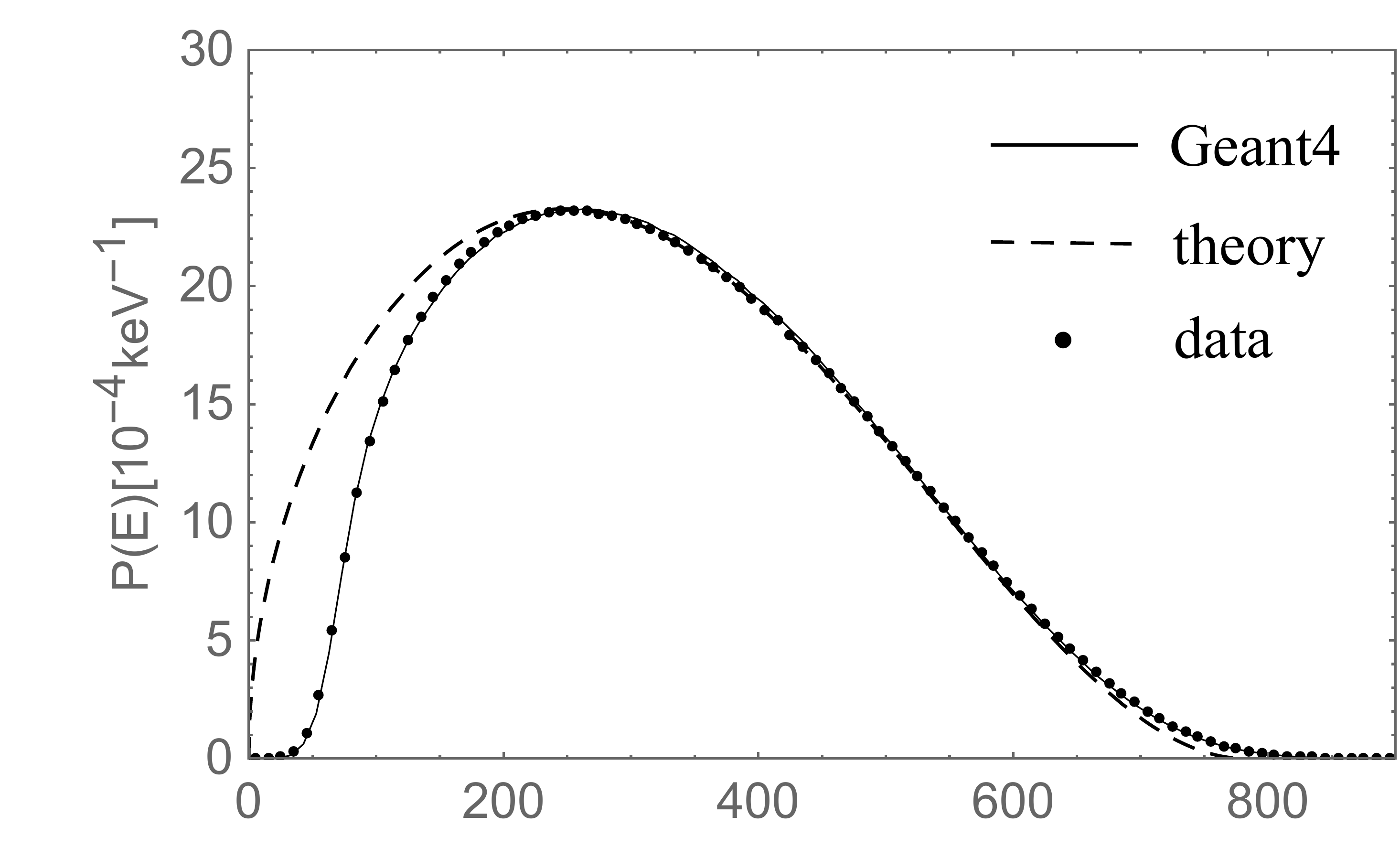}
            \vskip -0.18in
  \includegraphics[width=\linewidth]{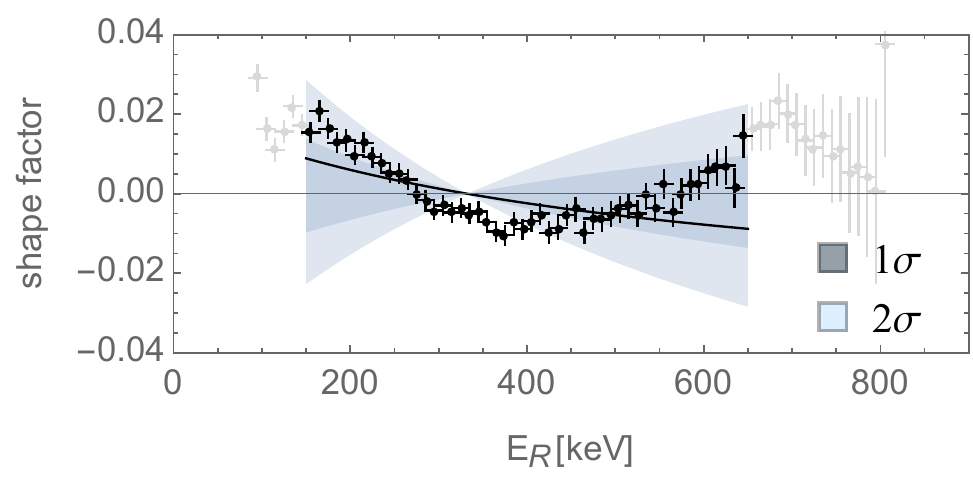}
  \vskip -.2in
	\caption {\label{fig:spectrum}
                Top: Measured energy spectrum compared with the full GEANT4 Monte Carlo spectrum for $b_n=0$ which includes the detector response. The dashed line is the predicted Standard Model spectrum in the absence of detector response. Bottom: Measured shape factor (as defined in text) with statistical uncertainties. The 1$\sigma$ and 2$\sigma$ uncertainty range due to energy response is also shown, based on the results from Fig. 3. The minimum in the uncertainty bands is due to the normalization of the fitted curves with varied energy response to the unvaried distribution.  
                }
\end{figure}
To assess the systematic effects for these varied energy responses, generated energy spectra from each varied response are fit, with $b_n$ as a free parameter, to the Monte Carlo spectrum without varied energy response and $b_n=0$. 
The results of the fitted values of $b_n$ for 500 simulated responses are shown in Fig. \ref{fig:bn-histogram}, with the quadratic polynomial assumption. From this distribution a systematic uncertainty due to energy response can be determined for both a $1\sigma\ (-0.056 \text{---} 0.087)$ and $90\%\ (-0.104 \text{---} 0.157)$ confidence interval. 

The experimental decay spectrum  $\Sigma(E)$ vs. reconstructed electron kinetic energy is shown in Fig. \ref{fig:spectrum} along with the full Monte Carlo spectrum for $b_n=0$. The expected spectrum from the Standard Model without energy response is shown as the dashed curve to indicate the effects of detector energy thresholds and energy resolution when compared to the measured spectrum. The lower panel in Fig. \ref{fig:spectrum} shows the shape factor, defined as 
$(\Sigma_M-\Sigma_\text{MC})/\Sigma_\text{MC}$, 
where $\Sigma_M$ is the measured spectrum and $\Sigma_\text{MC}$ is the simulated spectrum with $b_n=0$. A value for the Fierz interference term from the measured data is then determined by fitting the measured shape factor to that expected from Eq. \eqref{eq:bsm:Fierz-rate}. Since the systematic uncertainties increase significantly at both low and high energies (i.e. near detector threshold and $\beta$ spectrum endpoint) due to signal:noise degradation and energy response uncertainty, the spectrum is only fitted for electron kinetic energy $150 \text{ keV} < E < 650 \text{ keV}$. The fit to the data shows significant disagreement compared to the statistical uncertainties of the measurement likely because of the large systematic uncertainties. We note that the statistical uncertainties can be increased by a factor 2.5 to produce a reasonable fit, but this still leads to a statistical uncertainty much less than the systematic uncertainties discussed above.  

The best fit value is 
$b_n = 0.067 \pm 0.005_{\text{stat}} {{}^{+0.090}_{- 0.061}}{}_{\text{sys}}$,
where the systematic uncertainty is from the analysis discussed above. This corresponds to a 90\% confidence limit interval of $-0.041 < b_n < 0.225$ and, since the error on $\bF$ is much smaller, a similar limit for $\bGT$. The result is  consistent with a vanishing $b_n$ as predicted by the Standard Model and is dominated by the systematic uncertainty in energy response. This is the first direct extraction of $b_n$ from a measurement of the decay electron energy spectrum. Future  spectral measurements of Fierz interference (with proposed sensitivities $\leq 0.005$) will require significant improvements in characterization of the energy response of the detection system. Future improved sensitivity to $\bGT$ is also being investigated via a simultaneous fit to both the spectral and asymmetry energy dependence. 

This work is supported in part by the US Department
of Energy, Office of Nuclear Physics (DE-FG02-08ER41557, DE-SC0014622, DE-FG02-97ER41042),
National Science Foundation
(0653222, 
0700491, 
0855538, 
1002814,
1005233, 
1205977, 
1306997
1307426, 
1506459, and
1615153).
We gratefully acknowledge the support of the LDRD program (20110043DR), and the
LANSCE and AOT divisions of the Los Alamos National
Laboratory.

\bibliography{UCNA_b2010}
\end {document}